\documentstyle[12pt]{article}

\newcommand{\be}{\begin{equation}}
\newcommand{\ee}{\end{equation}}
\newcommand{\Dlt}{\Delta}

\newcommand{\prt}{\partial}

\newcommand{\al}{\alpha}
\newcommand{\ra}{\rightarrow}

\newcommand{\gm}{\gamma}

\begin{document}

\begin{center}

{\Large{\bf Self-similar power transforms in extrapolation problems} \\ [5mm]

S. Gluzman$^1$ and V.I. Yukalov$^{2,3}$} \\ [3mm]

{\it  $^1$Generation 5 Mathematical Technologies Inc.,
Corporate Headquaters, 515 Consumers Road, \\
Suite 600, Toronto, Ontario M2J 4Z2, Canada \\ [2mm]

$^2$Institut f\"ur Theoretische Physik, \\
Freie Universit\"at Berlin, Arnimallee 14, D-14195 Berlin, Germany \\ [2mm]
$^3$Bogolubov Laboratory of Theoretical Physics, \\ 
Joint Institute for Nuclear Research, Dubna 141980, Russia}

\end{center}

\vskip 0.5cm

\begin{abstract}

A method is suggested allowing for the improvement of accuracy of 
self-similar factor and root approximants, constructed from asymptotic
series. The method is based on performing a power transform of the given 
asymptotic series, with the power of this transformation being a control 
function. The latter is defined by a fixed-point condition, which improves 
the convergence of the sequence of the resulting approximants. The method
makes it possible to extrapolate the behaviour of a function, given as an 
expansion over a small variable, to the region of the large values of this 
variable. Several examples illustrate the effectiveness of the method.

\end{abstract}

\vspace{0.5cm}


\vskip 0.5cm

{\bf AMS classification numbers}: 40A05, 40A25, 40A30, 40G99, 40H05, 
41A05

\vskip 0.5cm

{\bf Corresponding author}:

\vskip 2mm

Prof. V.I. Yukalov

\vskip 1mm

Institut f\"ur Theoretische Physik, 

Freie Universit\"at Berlin, Arnimallee 14, D-14195 Berlin, Germany

\vskip 2mm

{\bf Tel}: 49 (30) 838-530-30 (office); 

\vskip 1mm

{\bf E-mail}: yukalov@physik.fu-berlin.de

\newpage

\section{Introduction}

In the majority of realistic computational problems, the sought function, 
satisfying a very complicated set of equations, cannot be defined for the 
whole range of its variable, but can be found only for asymptotically small 
values of this variable. At the same time, the most interesting could be 
the behaviour of the function at very large values of the variable. This 
is the standard situation in extrapolation problems, repeatedly appearing 
in various applications.

Suppose we are looking for a real function $f(x)$ of a real variable 
$x\in[0,\infty)$. By means of perturbation theory or an iterative procedure,
we can find the behaviour of this function at asymptotically small $x\ra 0$.
But what of the most practical interest in many cases is the behaviour of 
$f(x)$ at very large $x$, say as $x\ra\infty$. It is this the most difficult
extrapolation problem that we address in the present paper: How, knowing 
the behaviour of $f(x)$ only at $x\ra 0$, to define the value $f(\infty)$ for
$x\ra\infty$.

There exist several extrapolation methods, among which the most known are
the Pad\'e summation [1], Borel and Pad\'e-Borel summations [2], and the
optimized perturbation theory [3]. The latter was, first, advanced in 
Ref. [3] and nowadays is widely employed for various applications, as can 
be inferred from the review works [4,5]. Another extrapolation method is 
based on the self-similar approximation theory [6--11]. Using the techniques 
of this theory, supplemented by the fractal transforms [12,13], we have
recently derived a novel types of approximants allowing for an effective 
summation of power series and, hence, for their extrapolation. These are
the self-similar exponential approximants [14,15], self-similar root 
approximants [16--18], and  self-similar factor approximants [19,20]. 
The exponential approximants are appropriate for extrapolating functions
corresponding to exponentially varying processes, while the root and factor
approximants suit well for extrapolating functions with power-law behaviour. 
The self-similar approximants were shown to be simpler and more accurate 
than the Pad\'e approximants [16--20].

In the present paper, we aim at improving further the accuracy of the 
factor and root approximants by introducing a control function through 
a power transform of the initial asymptotic series. The new technique is 
illustrated by several examples of extrapolation from $f(x)$ at $x\ra 0$ 
to $f(\infty)$.

\section{Self-similar summation of power transforms}

Assume that the behaviour of the sought function $f(x)$ is known only 
for asymptotically small $x\ra 0$, when the function can be represented 
as an expansion in powers of $x$, being approximated by the series
\be
\label{1}
f_k(x) = \sum_{n=0}^k a_n x^n \; ,
\ee
where $k=0,1,2,\ldots$. Without the loss of generality, we may consider 
such expansions for which $a_0=1$, so that $f(0)=1$. Really, in the case
when
$$
f(x) \simeq f_0(x)\left ( 1 + a_1 x + a_2 x^2 + \ldots \right ) \; ,
$$
with a nontrivial $f_0(x)$, not expandable in a power series, we may 
always define 
$$
\overline f(x) \equiv \frac{f(x)}{f_0(x)} \; ,
$$
after which the series for $\overline f(x)$ acquire the form (1), where 
$a_0=1$. Our aim is, being based on the behaviour of $f(x)$ at $x\ra 0$, 
where it is approximated by the series (1), to find $f(\infty)$ at 
$x\ra\infty$.

The novel trick, we advocate in this paper, is to introduce a control 
function $m=m_k(x)$ by means of the {\it power transform}
\be
\label{2}
P_k(x,m) \equiv f_k^m (x) \; .
\ee
Taking the power $m$ of series (1), we reexpand the result in $x$ 
obtaining
\be
\label{3}
P_k(x,m) = \sum_{n=0}^k b_n(m) x^n \; ,
\ee
with $b_n(m)$ defined through $a_n$. A particular case of transform (2) is 
an inverted series with $m=-1$, which we have considered earlier. However,
fixing the power $m$ is not the best choice and here we shall advance a more 
general and rigorous way of selecting $m$. Expansion (3) serves as a basis 
for constructing in the standard ways [12--20] the self-similar factor and 
root approximants. Even-order factor approximants [19,20] are defined as
\be
\label{4}
F_{2k}(x,m) = \prod_{i=1}^k (1 + A_i x)^{n_i} 
\ee
and odd factor approximants can be represented as
\be
\label{5}
F_{2k+1}(x,m) = 1 + b_1 x \prod_{i=1}^k (1 + A_i x)^{n_i} \; ,
\ee
with the parameters $A_i=A_i(m)$ and $n_i=n_i(m)$ defined by the 
accuracy-through-order procedure with respect to series (3). This means 
that Eqs. (4) or (5) are to be expanded in powers of $x$, and these expansions
have to be compared with Eq. (3), equating the terms of like orders.

For the root approximants [16--18], we have
\be
\label{6}
R_{2k}(x,m) =\left ( \left ( \ldots (1 + A_1 x)^{n_1} + A_2 x^2
\right )^{n_2} + \ldots + A_k x^k\right )^{n_k}
\ee
in even orders, and
\be
\label{7}
R_{2k+1}(x,m) =1 + b_1 x \left ( \left ( \ldots (1 + A_1 x)^{n_1} + 
A_2 x^2 \right )^{n_2} + \ldots + A_k x^k\right )^{n_k}
\ee
in odd orders. The parameters $A_i=A_i(m)$ and $n_i=n_i(m)$ could be defined
in two ways. If the behaviour $f(x)$ at $x\ra\infty$ would be known, this 
could be used for uniquely defining all parameters [4]. Another way is to
determine these parameters by means of the accuracy-through-order procedure.
The second way may yield multiple solutions, which, however, are usually 
close to each other [21]. In what follows, we shall present only the most 
accurate approximant. The advantage of the root approximants is their 
ability to catch rather complicated asymptotic behaviour at large $x$, 
including corrections to the main scaling.

Defining the parameters $A_i(m)$ and $n_i(m)$ by the accuracy-through-order 
procedure, we assume that the limit $f(\infty)$ exists and finite, which 
imposes an additional constraint on the sum $\sum_{i=1}^k n_i$.

After constructing a factor approximant $F_k(x,m)$ or a root approximant 
$R_k(x,m)$ for the power transform (3), we have to accomplish the 
transformation inverse to Eq. (2), thus, obtaining either
\be
\label{8}
f_k(x,m) \equiv \left [ F_k(x,m)\right ]^{1/m}
\ee
or
\be
\label{9}
r_k(x,m) \equiv \left [ R_k(x,m)\right ]^{1/m} \; .
\ee

The improvement of the accuracy, as compared to the factor and root 
approximants not involving the power transformation (2), is achieved by
defining a control function $m=m_k(x)$ from a fixed-point condition. In 
general, there exist several types of such fixed-point conditions, which, 
actually, are equivalent to each other [4]. Here we use the simplest of
them, the minimal sensitivity condition, which gives either
\be
\label{10}
\frac{\prt}{\prt m} \; f_k(x,m) = 0 \; , \qquad m=m_k(x) \; ,
\ee
or
\be
\label{11}
\frac{\prt}{\prt m} \; r_k(x,m) = 0 \; , \qquad m=m_k(x) \; ,
\ee
depending on whether the factor approximant (8) or root approximant (9) is 
considered. With the so defined control function $m_k(x)$, we get either
\be
\label{12}
f_k^*(x) \equiv f_k(x,m_k(x))
\ee
or
\be
\label{13}
r_k^*(x) \equiv r_k(x,m_k(x)) \; .
\ee
From here, keeping in mind our main aim to extrapolate the sought function
to the limit $x\ra\infty$, we obtain either
\be
\label{14}
f_k^*(\infty) = \lim_{x\ra\infty}\; f_k^*(x)
\ee
or
\be
\label{15}
r_k^*(\infty) = \lim_{x\ra\infty}\; r_k^*(x) \; .
\ee
Below, we shall illustrate the method by several examples, confronting 
the found approximants $f_k^*(\infty)$ and $r_k^*(\infty)$ with known
values of $f(\infty)$. Note that for defining the limits (14) or (15),
we, actually, do not need to have the whole function $m_k(x)$, but what
we need to have is just a limiting value $m_k=m_k(\infty)$, which is a 
constant.

\section{Stirling series for factorial function}

Let us consider the factorial function
$$
f(x) = \frac{1}{\sqrt{2\pi}}\; e^{1/x} x^{1/x} \Gamma\left ( 1 +
\frac{1}{x}\right ) \; ,
$$
where $\Gamma(\cdot)$ is a gamma function. As $x$ tends to zero, one has
$$
f(x) \simeq \frac{1}{\sqrt{x}} \qquad (x\ra 0) \; .
$$
Therefore, we define the reduced function
$$
\overline f(x) \equiv \sqrt{x}\; f(x) \; ,
$$
whose small-$x$ expansion has the form of Eq. (1), so that, as $x\ra 0$,
then
$$
\overline f(x) \simeq 1 + a_1 x + a_2 x^2 + a_3 x^3 + a_4 x^4 + 
a_5 x^5 \; .
$$
The expansion coefficients are
$$
a_1 = \frac{1}{12}\; , \quad a_2 = \frac{1}{288} \; , \quad
a_3 = -\; \frac{139}{51840} \; , 
\quad a_4= -\; \frac{571}{2488320} \; , \quad
a_5 = \frac{163879}{209018880} \; .
$$
Thence, we shall apply the procedure of Section 2 to the function 
$\overline f(x)$, and at the end, we will return to the sought function 
$f(x)=\overline f(x)/\sqrt{x}$, looking for the limit $f(\infty)$. The
exact limit for the factorial is
$$
f(\infty) = \frac{1}{\sqrt{2\pi}} = 0.398942.
$$
Following the method, described in Sec. 2, we find $r_5^*(\infty)=0.458$, 
whose error, as compared with the exact $f(\infty)$, is $15\%$. The factor
approximant $f_5^*(\infty)=0.406$ is much better, with an error of only 
$2\%$. Comparing this with the Pad\'e approximants, we should remember that 
these are not uniquely defined, yielding for each given order a whole table 
of approximants [1]. One often considers solely the diagonal approximants. 
For the present example, the diagonal Pad\'e approximant $P_{[2/2]}$ describes
[22] the factorial-function limit $f(\infty)$ with an error of $14\%$. Thus, 
the factor approximant $f_5^*(\infty)=0.406$ is the most accurate. It is worth
emphasizing that a direct application of the factor-approximation technique, 
without involving the power transformation (2), would give the limiting value
$0.169$, which is a very bad approximation. Hence, employing the power 
transformation is a crucial point in improving the accuracy of the 
approximants.

\section{Debye-H\"uckel function for strong electrolytes}

The function
$$
D(x) = \frac{2}{x} \; - \; \frac{2}{x^2}\left ( 1  - e^{-x}\right )
$$
arises in the Debye-H\"uckel theory of strong electrolytes [23]. At small 
$x\ra 0$, this function possesses an expansion of the type (1), with
$$
a_1=-\; \frac{1}{3}\; \quad a_2 = \frac{1}{12}\; , \quad
a_3 = -\; \frac{1}{60}\; , \quad a_4 =\frac{1}{360} \; , \quad
a_5= -\; \frac{1}{2520} \; .
$$
We shall be interested in finding the limiting value $f(\infty)$ of the 
reduced function
$$
f(x) \equiv x D(x) \; ,
$$
whose exact limit is $f(\infty)=2$.

Using the technique of Section 2, we get for the best root approximant 
$r_5^*(\infty)=1.993$, whose error is $-0.4\%$. For the uniquely defined
factor approximant, we find $f_5^*(\infty)=1.779$, with an error of $-11\%$.
Note that without invoking the power transform, there are no real solutions 
for the sought limit. Thus, the usage of the power transformation (2) is 
principal here. The best Pad\'e approximant, employing the same coefficients
$a_n$, gives the limit $f(\infty)$ with an error of $-33\%$.

\section{Critical temperature of Bose gas}

Bose-Einstein condensation in dilute Bose gas has attracted much attention 
in recent years (see reviews [24--27]). One of the interesting problems,
which has been intensively studied, is the influence of atomic interactions 
on the shift of the critical temperature. One considers the relative 
variation of the critical temperature
$$
\frac{\Dlt T_c}{T_0}  \equiv \frac{T_c}{T_0}\; - 1 \; ,
$$
due to weak atomic interactions, as compared to the condensation temperature
$$
T_0 = \frac{2\pi\hbar^2}{mk_B}\left [ \frac{\rho}{\zeta(3/2)}
\right ]^{2/3}
$$
of the ideal homogeneous Bose gas. The lowest term in the expansion of 
the critical-temperature shift with respect to the small gas parameter
$$
\gm \equiv \rho^{1/3} a_s \; ,
$$
where $\rho$ is particle density and $a_s$, scattering length, is 
commonly represented as
$$
\frac{\Dlt T_c}{T_c} \simeq c_1 \gm \qquad (\gm\ra 0) \; .
$$
The coefficient $c_1$ has been calculated by a number of various methods.
Review of the related literature up to 2004 can be found in Refs. [25,27]. 
The most accurate are the results for $c_1$ obtained by means of the Monte
Carlo simulations and using the optimized perturbation theory. Less accurate
are the results based on a renormalization-group approach [28,29]. Lattice
Monte Carlo simulations by Arnold and Moore [30,31] give $c_1=1.32\pm 0.02$ 
and by Kashurnikov et al. [32,33], $c_1=1.29\pm 0.05$. Path integral Monte 
Carlo simulations by Nho and Landau [34] give $c_1=1.32\pm 0.14$. A variant 
[5] of optimized perturbation theory, employed by Kastening [35--37], yields
$c_1=1.27\pm 0.11$, and the optimized perturbation theory used by Kneur et 
al. [38,39], results in $c_1=1.30\pm 0.03$. Here, we shall calculate the 
coefficient $c_1$ by means of the technique of Section 2.

The coefficient $c_1$ can be expressed as an asymptotic expansion
$$
c_1(g) \simeq a_1 g + a_2 g^2 + a_3 g^3 + a_4 g^4 + a_5 g^5
$$
in powers of an effective coupling parameter [36], where
$$
a_1=0.223286\; , \qquad a_2=-0.0661032\; , \qquad 
a_3=0.026446\; , 
$$
$$
a_4=-0.0129177\; , \qquad a_5=0.00729073 \; .
$$
This expansion is valid for $g\ra 0$. But the sought value of $c_1$ is 
given by the limit
$$
c_1 = \lim_{g\ra\infty} c_1 (g) \; .
$$
Employing the factor approximants, complimented by the power transformation 
(2), we have $f_5^*(\infty)=c_1=1.09$, which is close to the values found 
by other methods. Summing the strong-coupling expansion with the help of 
the root approximants and defining the parameters from the weak-coupling 
expansion, we get $r_5^*(\infty)=1.19$.

\section{Structure factor of branched polymers}

The structure factor of three-dimensional branched polymers is given 
[40,41] by the confluent hypergeometric function
$$
S(x) = F_1\left ( 1; \frac{3}{2};\frac{3}{2}\;x\right ) \; .
$$
We shall consider the reduced function
$$
f(x) \equiv x S(x) \; ,
$$
whose limit $f(\infty)=1/3$ is finite. At asymptotically small $x\ra 0$, 
the function $f(x)$ possesses an expansion of the form (1), with 
$a_0=1$. Several other expansion coefficients are
$$
a_1=-1\; , \qquad a_2=0.6\; , \qquad a_3=-0.257\; , \qquad
a_4=0.086 \; .
$$
The best approximant, obtained by the method of Section 2, is 
$f_5^*(\infty)=0.329$, whose error is $-1.3\%$. This is much more accurate 
than the best Pad\'e approximant of the same order having an error of
$-266\%$.

\section{Discussion}

In this paper, we suggested a method for improving the accuracy of 
self-similar approximants by introducing a control function through the 
power transformation (2). As is shown by several examples, the accuracy 
really becomes essentially better.

Here we have concentrated our attention on the extreme extrapolation 
problem, when from the behaviour of a function $f(x)$ at asymptotically 
small $x\ra 0$ one has to find the limit $f(\infty)$ for $x\ra\infty$. 
This extrapolation problem is one of the most difficult. If we are able 
to accurately predict the behaviour of a function $f(x)$ at $x\ra\infty$,
then, as is clear, it is even easier to approximate its behaviour for 
finite $x$.

As an illustration of the latter statement, we may consider the expansion 
factor of a polymer. The properties of polymers are of great importance for 
a variety of applications [42]. Let us, for example, consider the expansion 
factor $\al(z)$ for a three-dimensional polymer chain with excluded-volume 
interaction, where $z$ is a dimensionless coupling parameter [43,44]. From
an asymptotic series of the type (1), derived by means of perturbation 
theory [43], we construct the expansion factor
$$
\al^*(z) = 1.5286 z^{0.3543} \left [ \left ( 1 + 0.1552 z^{-1}
\right )^{-0.0749} + 0.3302 z^{-0.9252} \right ]^{0.383} \; .
$$
This is obtained by considering a large-$z$ expansion, resumming it by means 
of the root approximants, and determining all unknown exponents and amplitudes
from the weak-coupling expansion of fourth order. The strong-coupling exponent
$$
\nu \equiv \frac{1}{2} + \frac{1}{4}\; \lim_{z\ra\infty}\;
\frac{\ln\al(z)}{\ln z}
$$
for our approximant $\al^*(z)$ is $\nu=0.5886$, which coincides with the
value found numerically [43--45]. That is, the approximant $\al^*(z)$
possesses a correct scaling behaviour. It also gives a nontrivial correction 
to the scaling, with an exponent of $-0.9552$. The expression $\al^*(z)$ is
valid for all $z\in[0,\infty)$, differing from the known numerical values 
[44] by not more than $0.3\%$.

\vskip 5mm

{\bf Acknowledgement}

\vskip 2mm

One of the authors (V.I.Y. ) is grateful to the German Research Foundation
for the Mercator Professorship.

\end{document}